\documentclass[11pt]{article}

\usepackage{epsfig}
\usepackage{amsfonts,amssymb,amsmath}

\newcommand{\sect}[1]{\setcounter{equation}{0}\section{#1}}
\renewcommand{\theequation}{\arabic{section}.\arabic{equation}}

\newcommand{\half}{\ensuremath{\frac{1}{2}}}

\newcommand{\Tr}{\ensuremath{{\rm Tr}}}

\newcommand{\be}{\begin{equation}}
\newcommand{\ee}{\end{equation}}

\newcommand{\ba}{\begin{eqnarray}}
\newcommand{\ea}{\end{eqnarray}}

\newcommand{\gym}{g_{\scriptscriptstyle \mathrm{YM}}}
\newcommand{\adss}{$ \mbox{AdS}_5 \times S^5 \, \mbox{} $}

\begin{document}

\bigskip

%
\hskip 4.8in\vbox{\baselineskip12pt
\hbox{hep-th/0403226}}

\bigskip 
\bigskip
\bigskip

\begin{center}
{\Large \bf Excited Mesons}\\
\bigskip
{\Large \bf and}\\
\bigskip
{\Large \bf Quantization of String Endpoints}
\end{center}

\bigskip
\bigskip
\bigskip

\centerline{\bf E. Schreiber}

\bigskip
\bigskip
\bigskip

\centerline{\it Department of Physics and Astronomy}
\centerline{\it University of British Columbia}
\centerline{\it 6224 Agricultural Road, Vancouver, BC, V6T 1Z1, Canada}
\vskip 6pt
\centerline{\it Pacific Institute for the Mathematical Sciences}
\centerline{\it University of British Columbia}
\centerline{\it 1933 West Mall,         Vancouver, BC, V6T 1Z2, Canada}
\vskip 6pt
\centerline{E-mail: \small \tt schreib@physics.ubc.ca}


\bigskip
\bigskip

\begin{abstract}
\vskip 2pt
\noindent
We argue that those field theories containing mesons that are dual to weakly curved string backgrounds are 
non--generic. 
The spectrum of highly excited mesons in confining field theories behaves as $M_n \propto \sqrt{n}$ 
(where $n$ is the radial excitation number), as does the spectrum of the dual mesons described by open 
strings ending on probe D branes. 
However, in the weakly coupled backgrounds, we show that the sector of (light) mesons with spin $J \le 1$, 
dual to the brane fluctuations, behaves as $M_n \propto n$, for confining and non--confining theories 
alike. 
In order to perform the analysis we suggest a method of quantization of the open string endpoints, 
and rely heavily on the semiclassical, or WKB, approximation.

\end{abstract}

\newpage

\baselineskip=18pt
\setcounter{footnote}{0}


\tableofcontents
\newpage

\sect{Introduction and Summary}
\subsection{Background and Motivation}

Understanding hadron physics, or the low energy, strong coupling regime of QCD, where confinement is 
manifest, is one of the most important unsolved questions in physics. Even finding from first principles 
the (discrete) spectrum would constitute a giant breakthrough.
In recent years, though, there has been a considerable advance, using holographic dualities of 
string theory (which usually go by the name of the AdS/CFT correspondence).   

The modern notion of a holographic duality between string theories and quantum field theories rests
on the existence of two complementary perspectives for the description of D branes. On the one hand,
they are solitonic objects in string theory and in its low energy limit, supergravity, which deform the 
background. On the other hand, the branes' worldvolume physics is, in a certain limit, a gauge quantum 
field theory \cite{Mal1}. 
In both descriptions it is crucial to use the defining property of the D branes, namely that strings 
might end on them. 

The simplest such duality, which was the first to be explored \cite{Mal1}, uses a stack of D3 branes.
On the one hand, this results in four--dimensional, ${\cal N} = 4$ supersymmetric, $SU(N)$ Yang--Mills 
theory (SYM) with gauge coupling $\gym$.
On the other hand, there is string theory on \adss, supported by $N$ units of flux; 
both factors have radius $R$ where $R^4 = 4 \pi g_s N \alpha'^2$. To suppress string interactions we prefer
taking $g_s \rightarrow 0$, while in order to suppress $\alpha'$ corrections to the supergravity limit of 
the string theory we concentrate on the regime with large $R$. As $g_s = 4 \pi \gym^2$, this results, for
the SYM side, in the 't Hooft limit of $N \rightarrow \infty$, $\gym \rightarrow 0$, with 
$\lambda \equiv \gym^2 N$ fixed and large.

The conformal ${\cal N} = 4$ SYM theory is still very different from the confining QCD theory. In order
to approach QCD, two things must be achieved: supersymmetry should be broken, and dynamical quarks should 
be added. A lot of progress has been made regarding the first of those issues, starting with Witten's
paper \cite{Witten:1998zw}, where compactified backgrounds dual to three-- and four--dimensional 
non--supersymmetric YM theories were presented. Those backgrounds, however, give rise also to Kaluza--Klein
modes which do not decouple from the YM theories.

Witten had argued \cite{Witten:argument} that this is a generic situation: 
a string background which is weakly curved (apart from a possible very small Kaluza--Klein factor)
will never give rise to YM decoupled from extra degrees of freedom. 
The reasoning is very simple. A weakly curved background is well approximated by supergravity, 
containing only the massless closed string modes of spins $J \leq 2$. The higher spin string modes are 
very massive and therefore qualitatively different. However, string modes of all spins are presumably 
dual to glueballs, and in YM there is no sharp distinction between the $J \leq 2$ ones and the higher 
spin ones --- they are all supposed to lie on the same Regge trajectories.

Returning to the SYM scenario, we may approach the addition of quarks by separating one D3 brane from
the stack. This separation is equivalent to the Higgsing of the gauge group from $SU(N)$ to 
$SU(N-1) \times U(1)$, and a string from the stack to the separated brane is a W$^\pm$ particle.
From the point of view of the $SU(N-1)$ theory, though, it is something of a quark. Such a pair of quark 
and anti--quark may lower its energy by merging the endpoints of the strings on the stack and forming a 
bound state which is a smooth string. This enables us to calculate the potential between the quark and 
anti--quark \cite{Mal2, ReYe}. The theory is conformal, so the potential must be inversely proportional to 
the separation $L$ between the quarks; the fascinating result is that large screening is manifested in 
this strongly coupled regime: $E \propto - \frac{\sqrt{\lambda}}{L}$ instead of the perturbative
$E \propto - \frac{\lambda}{L}$. This is the classical result, but the corrections arising from quantum
fluctuations have relative strength of $\frac{\alpha'}{R^2} \sim \frac{1}{\sqrt{\lambda}}$ so can be 
safely ignored.

When the background is less supersymmetric than \adss, it might lead to confinement 
in the dual field theory \cite{Witten:1998zw, BISY2}.

However, for this picture to really make sense, the additional D brane must be infinitely far away, 
and the quarks must be infinitely heavy, or external; otherwise, the quark and anti--quark will attract 
each other until they meet and perhaps annihilate. There is no static classically stable configuration 
when the additional brane is only finitely removed. The string can be stabilized, though, by the 
centrifugal force when it is given angular momentum $J$ (corresponding to the meson having that spin). 
When $J$ is large, the classical string configuration is a good approximation \cite{Gubser:2002tv}. 

Recently, a way was found to add dynamical quarks to the AdS/CFT correspondence, that is, 
to incorporate finite mass particles in the fundamental representation of the gauge group.
This is achieved by the inclusion of appropriate D7 brane probes, in addition to the D3 branes 
responsible for the usual correspondence \cite{Karch:2002sh, Karch:2002xe}. The backreaction of the D7
brane on the background is small and will be neglected in the following.  
Similar strategies involving higher dimensional D branes apply to the deformed backgrounds discussed above.
High spin mesons can be described again as rotating strings. Moreover, the string dual to the meson 
can also be stabilized by rotation in the extra dimensions of the brane; 
this corresponds to the meson having (high) R--charge $J_R$ 
\cite{Berenstein:2002jq, Gubser:2002tv}.
 
The spectrum of mesons with low spins, $J \le 1$, can be derived from the spectrum of fluctuations of the 
higher dimensional D brane in the background in which it is embedded \cite{Kruczenski:2003be}. 
This is so as the higher dimensional D brane worldvolume theory encodes the dynamics of the massless 
open string degrees of freedom living on it --- the bosonic ones being scalars and a vector. 
We stress that all excited states with the same (low) spin can be found. 
Some further works deal with cases having less supersymmetry and will be referred to in the following. 

\subsection{Results and Methods}

As we have seen, there are two kinds of mesons in theories dual to superstring theory on weakly curved 
background: the lower spin ones, with $J \leq 1$, dual to D brane fluctuations, and the higher spin ones, 
with $J > 1$, described by strings ending on those D branes (the conditions for this string description to
be good will be discussed below). We claim that {\em those lower spin mesons
behave very differently than their counterparts in generic field theories such as QCD}. In a sense, our
argument is the open string analogue of Witten's argument referred to in the above. We will be forced to 
conclude that the field theories we are able to study using the AdS/CFT duality and its generalizations
are non--generic also in the behaviour of their meson sector.

The first difference has already been noticed in the analyses of 
\cite{Kruczenski:2003be, Kruczenski:2003uq, Hong:2003jm}, namely, that those mesons tend to be anomalously 
light. The mass of the two quarks is almost completely annihilated by the attracting potential,
resulting with an almost massless bound state. Let us look for example at the simplest case of quarks
added to the conformal ${\cal N} = 4$ SYM theory. The only scale of the theory then is the mass $m_q$
of the bare quark. The mass of the scalar meson in its ground state, however, turns out 
\cite{Kruczenski:2003be} to behave as $M_1 \sim \lambda^{-1/2} m_q$. As the 't Hooft coupling $\lambda$
is taken to be large, this is indeed very small.

We argue that the difference is also seen in the dependence of the meson masses $M_n$ on the radial
excitation number $n$, for a given value of the spin and R--charge. We claim that the generic confining 
gauge field theory result, for large $n$, is $M_n \propto \sqrt{n}$. This has been shown analytically 
in the case of two dimensional QCD in the large $N$ limit \cite{'tHooft:1974hx}. Moreover, this behaviour 
follows from general sum rule results coming from the dispersion relation 
\cite{Shifman:bx, Zhitnitsky:1985um, Kogan:1995nd}: see appendix \ref{Appendix:DispersionRelations} for a 
review.

This behaviour, we argue further, is reproduced by the picture of mesons as open strings ending
on the probe D brane in the appropriate background dual to the confining gauge theory. However, we show
that from the picture of low spin mesons as fluctuations of the probe D brane one generically obtains a 
very different behaviour, namely $M_n \propto n$.

Solving exactly the equations governing the string suspended from the probe brane (dual to the high spin 
meson) is obviously a very hard problem. However, we have seen that the brane configuration can supply a 
potential for a given position of the string ends. 
From quantum mechanics we are used to the idea that the wavefunction is smeared around the minimum
of the potential. The position of the string endpoints on the removed brane give the positions of 
the quarks in the field theory (although there is a smearing due to the fact that the brane is in a finite
radial position rather than at the boundary). We therefore propose that
{\em the energy of the configuration should be computed for the positions (and momenta) of the string 
endpoints; this energy should then be taken as the Hamiltonian for the quantized quarks wavefunction.}
The energy of the configuration depends, of course, on the details of the interpolating string. 
There is, however, an effective potential resulting from the relaxation of the string. We further propose
that {\em in certain regimes, the string can be taken to have the minimal energy, that is, it should obey 
the classical equations of motion and the quantum fluctuations can be ignored}. 

This suggestion is reminiscent of the Born--Oppenheimer approximation in molecular physics. 
The nuclei are the ``slow'' degrees of freedom (corresponding to the quarks), while the electrons are the 
``fast'' degrees of freedom (corresponding to the string). For a given position of the nuclei,
the electrons are ``integrated out'', by solving their Schr\"odinger equation. The resulting energy is the
emerging effective potential for the nuclei. Subsequently, the nuclei wavefunctions and the system's total
energy are obtained by quantizing the nuclei (solving their Schr\"odinger equation) in the effective 
potential. Notice that in a sense our suggestion is even simpler, as the effective potential is found from 
the classical, not the quantum, behaviour of the ``fast'' degrees of freedom. However, notice also that
in the atomic analogue the ``slow'' degrees of freedom are the heavy ones, and the ``fast'' are 
the light ones, while in our cases the roles are interchanged.

We further note that in order for the classical picture of the string to hold, and for our suggestion 
to make sense, {\em the string should be much longer than its natural length scale}. 
This can be seen in several ways. 
For the D brane models, only when the length is large is the string having a noticeable dip into the bulk.
Otherwise, the string lies flat near the brane, and behaves like the usual free string in flat Minkowski 
space, which is highly non--classical. 
In order to have a long string, {\em the system should have large quantum numbers}. 
One possibility is having large angular momentum, supplying a centrifugal potential.
The classical behaviour of strings in that regime has already been paid attention to lately, 
following \cite{Gubser:2002tv}. We can use the scenario of quarks added to the conformal ${\cal N} = 4$
SYM theory to lend more credibility to our suggestion. The meson in this case, better described as 
positronium, was found \cite{Kruczenski:2003be} to behave indeed, for $J \gg \sqrt{\lambda} \gg 1$,
 as a pair of quark and anti--quark bound by the Coulombic potential arising from the classical string.  
Our suggestion, however, is less interesting in the stationary case of a rotating string of classically 
fixed radius. The reason is that this results with the quark's wavefunction being nothing more than a thin 
density shell around the classical value. 

Another possibility for having a long string is it possessing a large R--charge $J_R$, for this is 
nothing but angular momentum in the extra dimensions of the brane.  

We will concentrate, however, on the third possibility, namely, having a high (radial) excitation
number $n$.

For a linearly confining theory we claim that we may safely use a na\"{\i}ve model, which 
nevertheless supplies us with important results and intuition. We simply take
a string in flat Minkowski space with two particles attached at its end (those particles may well be 
D0 branes, but we will refer to them as quarks, although we treat them as spinless and as having no 
charge). This is similar, for heavy quarks, to the well known 
treatment of bag models, but in those cases, like the Charmonium spectrum and wave function determination,
the confining potential is extracted in a phenomenological manner. In appendix \ref{Appendix:YM3} we 
argue for the validity of the approximation using as an example Witten's three--dimensional Yang--Mills
scenario \cite{Witten:1998zw}. 

This na\"{\i}ve model has been explored already in the early days of string theory 
(see, for example, \cite{Bardeen:1975gx, Ida:1977uy, LaCourse:1988cu, Allen:2001wu} 
and references therein).
The justification for treating the string as classical, although its energy is dominant (it is the 
``heavy'' degree of freedom) rests on the fact that it is straight in all of our applications 
\cite{Ida:1977uy}. We might then say that it is ``easy'' for the string to follow its endpoints.
Furthermore, in this model, one sees indeed that the necessary and sufficient
condition for the string to be nearly classical is it being long. Formally, the L\"uscher term 
\cite{Luscher}, which is the lowest order (quadratic) contribution of quantum fluctuations, 
is then negligible.
We review this well known story for the static string in appendix \ref{Appendix:Luscher}, 
and for the stationary rotating string in appendix \ref{Appendix:RotatingString}. 
It is also easy now to see intuitively
the claimed relation of the mass and excitation number, $M_n \propto \sqrt{n}$. In the highly excited 
state, the quarks at the string endpoints are less important; ignoring them, we know that the mass squared
grows like the discrete excitation number (which might manifest itself either as radial excitation or 
as angular momentum).

It is important to understand that the two radial excitations we have described occur in different spaces:
the field theory space for the string excitations, and the extra directions of the probe brane for the
case of its fluctuations. Strictly speaking, we should have called the excitation number in the latter 
case $n_R$. However, the spectra we have described are the lowest energy excitations in both cases.
In the brane fluctuation picture, excitations in the field theory space can not be described.
Such excitations should be described by the other, much heavier mechanisms. In the suspended string  
picture, on the other hand, excitations in the extra directions are possible. However, as in such a case
the radial coordinate $U$, which sets the energy scale, increases, simple monotonicity arguments show that
the energies are larger than the corresponding excitations in the field theory space, where $U$ is 
constant. We also present an explicit calculation verifying this claim in appendix \ref{Appendix:x4string}.

We wish to address a further point here. The quark anti--quark pair might annihilate; indeed, mesons have 
a finite lifetime. However, this corresponds to the two string endpoints merging, which is an $O(g_s)$ 
effect. As $\lambda$ is kept fixed (though large) while $N$ is taken to infinity, this effect is also 
negligible. The exact value of $g_s$ is also responsible for the 
D brane tensions (in particular, the D0 mass), but this will be of no importance to us. 

Solving the system according to our suggestion is still not an easy task. Happily, for the highly excited
meson we might resort to the semiclassical, or WKB, approximation. This was suggested for the na\"{\i}ve
model already in \cite{Bardeen:1975gx, Ida:1977uy}.

\subsection{Organization of the Paper}

The semiclassical (WKB) method is reviewed extensively in section \ref{WKB}. 
First we describe it for the non--relativistic Schr\"odinger equation (this should be rather familiar, 
and can be only glanced at). 
The highly excited states, though, are necessarily relativistic, so we next review the very similar case 
of the Klein--Gordon equation. 
Then we describe the aforementioned general monotonicity properties in the WKB approach.
It turns out that the WKB method is also applicable to ``Laplacian type'' 
equations arising from the study of the spectrum of probe D brane fluctuations, which are dual to the low 
spin mesons! This is also explained in section \ref{WKB}; we defer the presentation of an alternative
viewpoint on the relation $M_n \propto n$ in these cases to appendix \ref{Appendix:MoreLaplacian}.

In section \ref{Mesons:SpinningStrings} we apply our suggestion for the quantization of the string 
endpoints, using the WKB approximation. We concentrate on the na\"{\i}ve model of linear confinement,
taking the quarks' masses and the angular momentum of the system into consideration. 
We stress that even though those strings are spinning (they have $J > 1$, and possibly $J \gg 1$), 
the cardinal excitation is the radial one, as $n \rightarrow \infty$. 
We indeed find that the meson spectrum dual to those strings behave as $M_n \propto \sqrt{n}$. 

In section \ref{Mesons:BraneFlucts} we explore the low spin meson spectrum dual to fluctuating brane 
probes in five different backgrounds: the conformal \adss one, and four confining ones,
the Klebanov--Strassler geometry, Witten's geometry dual to four--dimensional Yang--Mills theory,
the Constable--Myers geometry and the Maldacena--N\'u\~nez geometry.
We apply the WKB approximation to prior analyses and find in all cases that for high excitations, 
the spectrum behaves as $M_n \propto n$.

A discussion follows, where we outline the open questions and future directions. 
We end with the six appendices described above.

We work in units where $\hbar = c = 1$, but sometimes keep $\hbar, c$ explicit.


\sect{The Semiclassical (WKB) Approximation}
\label{WKB}

In this section we first describe the well known semiclassical, or WKB, approximation for the Schr\"odinger
equation. We show next that the method is virtually the same for the relativistic Klein--Gordon equation,
which is relevant, by our string endpoint quantization suggestion, to the mesons of spin $J > 1$. 
We then explore the monotonicity properties of the method.
Finally we show that some ``Laplacian type'' equations, governing the probe D brane fluctuations 
(and therefore relevant for the mesons of low spin, $J \le 1$), can be transformed to the Schr\"odinger 
equation, and therefore also analyzed by the WKB method. 

\subsection{The Schr\"odinger Equation}

The Schr\"odinger equation in one dimension,
\be
\label{SchrUsual}
- \frac{\hbar^2}{2 m} \partial_x^2 \psi(x) + V(x) \psi(x) = E \psi(x)
\ee
is the equation attained from the classical non--relativistic expression for the energy of a particle of
mass $m$ in a potential $V(x)$, that is, from
\be
\label{ClassicalE}
\frac{p^2}{2m} + V(x) = E
\ee 
when $p$ is interpreted as the operator $-i \hbar \partial_x$.
the basic problem is to find the discrete spectra, $E_n$, and wavefunctions, $\psi_n(x)$, of the states 
bound to the potential $V(x)$, with $n = 1,2,\ldots$. Those states are orthonormal,
\be
\label{SchrUsualNorm}
\int \psi^*_n(x) \psi_m(x) dx = \delta_{n m}
\ee
The semiclassical (or WKB) method is an approximation valid for large $n$. Let us define the
classical momentum, which is the solution of (\ref{ClassicalE}), as
\be
p(x) = \sqrt{2 m E_K} = \sqrt{2 m (E - V(x))}
\ee   
and the turning points $x_-,x_+$ as the points where classically the particle stops and turns around,
$V(x_\pm) = E$ or $p(x_\pm) = 0$.
Let us also rewrite the wavefunction as 
\be
\label{psisigma}
\psi(x) \equiv \exp\left(\frac{i}{\hbar} \sigma(x)\right)
\ee
and expand $\sigma(x)$ in powers of $\hbar$,
\be
\sigma(x) = \sigma_0(x) + \hbar \sigma_1(x) + \ldots
\ee
As 
\be
\psi''(x) = \left[\frac{i}{\hbar} \sigma''(x) +  \left(\frac{i}{\hbar} \sigma'(x)\right)^2\right] \psi(x)
\ee
the Schr\"odinger equation (\ref{SchrUsual}) turns out to be
\be
\left(\sigma'(x)\right)^2 - i \hbar \sigma''(x) = p^2(x)
\ee
If we keep only the leading order equation in $\hbar$ we get a first order differential equation
\be
\left(\sigma_0'(x)\right)^2 = p^2(x)
\ee
having the immediate solutions
\be
\sigma_0(x) = \pm \int^x p(\tilde{x}) d\tilde{x}
\ee
Actually, in one dimension the wavefunction can be taken as real, so we have from 
(\ref{psisigma}) that 
\be
\label{psi0}
\psi(x) \approx \sin \left(\frac{1}{\hbar} \int^x       p(\tilde{x}) d\tilde{x}\right)
\ee
where the lower integration point, or equivalently an additional phase in the sine function, 
is not yet determined. Note also that this wavefunction is not normalized.
We now make the further tentative approximation that the wavefunction vanishes at the turning points 
$x_\pm$. This is true for an infinite square potential well, and although there is tunneling in other 
cases, the wavefunction diminishes exponentially in the classically inaccessible region near the turning 
points, so the inaccuracy is not that large. This assumption allows us to take the aforementioned 
integration point as $x_-$. Then, in order for the wavefunction to vanish also at $x = x_+$, one needs
\be
\label{En0}
S \equiv \int_{x_-}^{x_+} p(x) dx \approx \pi \hbar n
\ee
where $n$ is integer. $S$ is the classical action of one period of the motion, and is an adiabatic 
and canonical transformation invariant. We have arrived at the Bohr--Sommerfeld quantization condition; 
this semiclassical condition implicitly gives $E_n$ as a function of $n$.

It is not too hard to refine the analysis by finding the next order approximation $\sigma_1(x)$ 
and by matching more carefully the true behaviour of the wavefunction near $x = x_\pm$, that is, 
taking more accurate boundary conditions. One then finds that the $n$ in (\ref{En0}) is replaced by
$n + \half$, which is of little interest to us\footnote{then, $n=0$ is also allowed. We, however, will
keep the convention $n \ge 1$}, 
and that (\ref{psi0}) is modified such that the wavefunction $\psi_n(x)$ is approximated at 
$x_- \ll x \ll x_+$ by
\be
\label{psi1}
\psi(x) \approx \frac{1}{\sqrt{p(x)}} \; 
                \sin \left(\frac{1}{\hbar} \int_{x_-}^x p(\tilde{x}) d\tilde{x}\right) 
\ee

This modification is fortunate for the following reason.
In our applications we cannot actually trust the potential $V(x)$ at its center, or minimum:
the na\"{\i}ve model is no longer applicable, and quantum effects are important. However, for 
high enough energies this doesn't matter very much, as this only changes $p(x)$ by a small amount over a 
fixed range. Also, the wavefunction is small in this dubious range of $x$, as, by (\ref{psi1}),
$|\psi(x)|^2 \propto \frac{1}{p(x)}$ is small there. In fact, this has a simple classical analogue, 
as required by the correspondence principle: because $p(x)$ is proportional to the classical speed of
the particle, the probability to find the particle between $x$ and $x+dx$, which is $|\psi(x)|^2 dx$,
 is proportional to the time $dt$ the particle stays there. 
 
\subsection{The Klein--Gordon Equation}

The Klein--Gordon equation stems from the relativistic expression of a particle's energy, still in a 
potential well, 
\be
\sqrt{p^2 + m^2} + V(x) = E 
\ee 
or
\be
\label{Relativisticp2}
p^2 = \left(E - V(x)\right)^2 - m^2
\ee
with the same replacement $p \rightarrow -i \hbar \partial_x$. Explicitly, it reads
\be
- \hbar^2 \partial_x^2 \psi(x) = \left[\left(E - V(x)\right)^2 - m^2\right] \psi(x)
\ee
and is sometimes denoted as the ``spinless Salpeter equation''.
Exactly the same procedure can be taken for finding the semiclassical approximation as in the case of the
Schr\"odinger equation. At the leading order in $\hbar$ one still gets (\ref{En0}), but where now 
the appropriate classical momentum is, from (\ref{Relativisticp2}),
\be
p(x) = \sqrt{\left(E - V(x)\right)^2 - m^2}
\ee
This will be all we need for the applications of this paper.
 
\subsection{Monotonicity Properties}
\label{WKB:Monotonicity}
Let us now compare two different potentials, either in the Schr\"odinger or the Klein--Gordon case,
such that for all $x$, $V^{(1)}(x) \ge V^{(2)}(x)$. For a given energy $E$, the corresponding classical 
momenta clearly obey $p^{(1)}(x) \le p^{(2)}(x)$, and the turning points, 
$x_-^{(2)} \le x_-^{(1)} < x_+^{(1)} \le x_+^{(2)}$. Consequently for the actions, 
$S^{(1)}(E) \le S^{(2)}(E)$ and therefore for all $n$ one has $E^{(1)}_n \ge E^{(2)}_n$ as expected.

The same argument shows that for different masses, $m^{(1)} > m^{(2)}$, again $E^{(1)}_n \ge E^{(2)}_n$ 
in the Klein--Gordon case. For the Schr\"odinger case, one needs to add the rest energy $m c^2$ to the 
Hamiltonian in order to get this result.

As we have only used the monotonicity of the energy in $V(x)$ and in $m$, those results are clearly more
general.

\subsection{Laplacian Type Equations}
\label{WKB:Laplacian}

We also encounter a different kind of second order differential equation, arising from eigenvalue problems
of the Laplacian on curved manifolds. For a space with Lorentzian metric
\be
ds^2 = g_{M N} dx^M dx^N,
\ee
the Laplacian, or rather the d'Alembertian, is 
\be
\label{GeneralLaplacian}
\Delta = -\frac{1}{\sqrt{|\det g|}} \partial_M \left(\sqrt{|\det g|} g^{M N} \partial_N\right)   
\ee  
In the cases of probe D brane actions, the space is a (warped) product of the Minkowski space, with 
coordinates $x^\mu$, where the 
field theory is defined, and another, Riemannian space. This Riemannian space can usually be written in 
terms of a radial coordinate, which we will denote by $y$, and a sphere. The metric is then of the form
\be
ds^2 = F_y^2(y) \, dy^2 + F_s^2(y) \, d\Omega^2 + F_M^2(y) \, (\eta_{\mu \nu} \, dx^\mu dx^\nu)
\ee
The eigenfunction $\Psi$ of the full Laplacian (\ref{GeneralLaplacian}) is then a product of a plane wave 
$\exp \left(i \, \eta_{\mu \nu} \, x^\mu k^\nu\right)$ in Minkowski space, a spherical harmonic on the 
sphere, and a function $\chi(y)$.
The eigenvalue of the Laplacian on the Minkowski space is $k^2 = -M^2$ where $M$ is the mass of the 
corresponding meson, and the eigenvalue of the Laplacian on the sphere is also known in terms of the
spherical harmonic, or of the R--charge $J_R$.
Then, the wave equation $\Delta \Psi = 0$ simplifies considerably and becomes an ordinary differential 
equation with coordinate $y$. 
Changing the notation for future convenience, and possibly redefining $\chi(y)$ properly, 
the eigenvalue problem is to find
the eigenfunctions $\chi_n(y)$ and the eigenvalues $\lambda_n$ such that\footnote{$\Gamma(y)$ should not be
confused, of course, with Euler's Gamma function. Also, $\lambda$ here is an eigenvalue and not the 
't Hooft coupling.}
\be
\label{SchrLap}
-\frac{1}{\Gamma(y)} \partial_y \left[ \frac{\Gamma(y)}{\Sigma^2(y)} \partial_y \chi(y) \right] 
 + A(y) \chi(y) = 
\lambda \chi(y)
\ee
where $\lambda = M^2$, and where $A(y)$ arises from the spherical harmonic part. 

The normalization condition now is 
\be
\label{SchrLapNorm}
\int \Gamma(y)       \chi^*_n(y) \chi_m(y) dy \equiv
\int \sqrt{|\det g|} \chi^*_n(y) \chi_m(y) dy =      \delta_{n m}
\ee
The variable $y$ might have an infinite range $-\infty < y < \infty$, where usually there is a reflection 
symmetry $y \rightarrow -y$. Most times, though, $y$ is a radial coordinate, as we have argued, 
having the range $0 \le y < \infty$. However, in those cases the boundary condition at 
$y = 0$ is usually either Dirichlet, $\chi(0)=0$, or Neumann, $\chi'(0)=0$, and the wavefunction can
be continued to an odd or even one, respectively, on the whole range $-\infty < y < \infty$.
 
Those Laplacian eigenvalue problems are relevant to our study as they occur in the quadratic fluctuation 
approximation to (Chern--Simons)--Born--Infeld brane actions, therefore determining the spectrum of mesons
with $J \le 1$.  
They are akin to the relativistic Klein--Gordon equation and not to the non--relativistic Schr\"odinger
one; in particular, $\lambda$ is a measure of the mass (or energy) 
{\em squared}. However, we can formally convert (\ref{SchrLap}), (\ref{SchrLapNorm}) into 
(\ref{SchrUsual}), (\ref{SchrUsualNorm}). We will further assume without loss of generality that 
$\hbar = 1$ and $m = \half$, so that the Schr\"odinger equation will be simply
\be
\label{SchrSimp}
- \psi''(x) + V(x) \psi(x) = E \psi(x)
\ee

The conversion is achieved by relating the coordinates,
\ba
\frac{d x}{d y} & = & \Sigma(y) \\
\label{SchrLapxy}
x               & = & \int^y \Sigma(\tilde{y}) d\tilde{y} 
\ea 
and the wavefunctions,
\be
\psi(x) = \Xi(x) \chi(y(x))
\ee
where
\be
\Xi(x) \equiv \sqrt{\frac{\Gamma(y(x))}{\Sigma(y(x))}}
\ee
The resulting potential in the Schr\"odinger equation is 
\be
\label{VxXiA}
V(x) = \frac{\Xi''(x)}{\Xi(x)} + A(x)
\ee
where the functions are now thought of as depending on $x$.
We further need to identify $E = \lambda$. 

The WKB method can be now applied to the Schr\"odinger equation corresponding to (\ref{SchrLap}) in order
to find the asymptotic behaviour of $\lambda$ as a function of $n$. The generic case is to find that
even if the range of $y$ is infinite, we get a finite range of $x$. From (\ref{SchrLapxy}), this occurs 
when 
\be
\label{x0}
x_0 \equiv \half \int_{-\infty}^{\infty} \Sigma(y) dy < \infty
\ee
Then, without loss of generality we may choose the range $-x_0 < x < x_0$. Typically in the reflection
symmetric case, $\Xi(x)$ behaves as a power at $|x| \lesssim x_0$, that is, 
$\Xi(x) \sim \left(x_0 - |x|\right)^a$ for some $a$. Let us assume for the moment that $A(y) \equiv 0$.
When $a > 1$ or $a < 0$, we get that the potential behaves near $\pm x_0$ as 
$V(x) \sim +\left(x_0 - |x|\right)^{-2}$, and in particular diverges, 
$V(\pm x_0) = +\infty$. Usually, the addition of a non--zero $A(y)$ does not change this divergence.
In that case for large energies we get essentially an infinite square potential well with turning points 
$x_\pm \approx \pm x_0$, and so, in our conventions, the energies behave as 
$E_n \approx \frac{\pi^2}{4 x_0^2} n^2$. As the energies equal the 
Laplacian eigenvalues which correspond to the masses squared, we get that generically the masses of the 
$J \le 1$ mesons involved are linear in $n$,
\be
\label{Mn}
M_n \approx \frac{\pi}{2 x_0} n
\ee

The analysis shows that the result is very generic, stemming only from the finite range of $x$. 
Indeed it seems to follow just from the relativistic behaviour of the constituents of the mesons 
at very high levels.


\sect{Spectrum of Mesons Dual to Excited Strings}
\label{Mesons:SpinningStrings}

In this section we explore the spectrum of mesons with spin $J > 1$ in theories having weakly curved
dual supergravity backgrounds. The spin $J$ should be thought of as fixed, while the radial excitation 
number $n$ tends to infinity, so the dynamics are dominated by it. 
The mesons themselves are dual to spinning strings whose endpoints are
stuck on static brane probes placed in those backgrounds. We find the robust result that in linearly 
confining theories, the spectrum behaves as $M_n \approx M_1 \sqrt{n}$, where $n$ is the excitation number.
For other quark anti--quark potentials the dependence is markedly different. In particular, the bound 
states in non--confining theories have negative energies which tend to zero as $n$ grows; this is the 
same qualitative behaviour as in the (non--relativistic) hydrogen atom problem.

We begin by looking at two relativistic spinless and chargeless particles (the ``quarks''), 
which are massless for the time being, placed at positions $x, -x$. 
The two particles are connected by a ``string'' giving rise to a potential which we take of the form 
\be
V(x) = 2 a |x|^\alpha
\ee
For the time being we assume that the particles have only radial and no transverse motion, so that the
spin of the system is $J = 0$. Although this physical picture applies for mesons with $J > 1$, we can
still consider $J$ to be small and ignore it for large energies. We will substantiate this claim better
later on. 

We are ignoring the center of mass degree of freedom as it decouples, 
and so we may consider only one particle. 
Without loss of generality we might choose the right quadrant of the motion, having $x \ge 0, \, p \ge 0$,
so we may assume that the energy of (half) the system is given by
\be
E = p + a x^\alpha
\ee
The massless particles move at the speed of light, and change direction abruptly when they hit the wall
of the potential well. Actually, the system without the quark's degrees of freedom does not make any sense,
as a Hamiltonian of the form $H = V(x)$ does not allow any motion. In a sense, the quarks absorb whatever 
energy needed from the string.
The semiclassical treatment of the Klein--Gordon equation can now be employed.
The turning point is $x_+ = \left(\frac{E}{a}\right)^{1/\alpha}$ and so the quantization condition
(\ref{En0}) is applicable with $p(x) = E - a x^\alpha$. Looking first at the confining theories
with $\alpha > 0, a > 0$, we find
\be
\label{nalpha}
n \hbar \approx 4 \int_0^{x_+} p(x) dx = 4 \frac{\alpha}{\alpha+1} E^{(\alpha+1)/\alpha} a^{-1/\alpha} 
\ee 
so that, approximately,
\be
\label{Enalpha}
E_n \propto n^{\alpha/(\alpha+1)}
\ee

We indeed see that linear confinement, that is $\alpha = 1$, leads to $E_n \propto \sqrt{n}$, where $E_n$
should be interpreted as the mass $M_n$ of the $n$-th excited meson.
Having $E_n \propto n$ would necessitate, on the other hand, sending $\alpha$ to infinity, resulting in 
the square well potential $V(x) = \begin{cases} 0 & |x| < x_+ \\ \infty & |x| > x_+ \end{cases}$, 
which is clearly not physical --- corresponding, so to speak, to a rope rather than to a string.

Negative $\alpha$ and $a$ give rise to non--confining potentials.
The range $\alpha \le -1$, however, gives $p(x)$ too strong a singularity at $x = 0$ to be a 
sensible model; an effective cut--off can be introduced if the angular momentum $J$ is taken as non--zero. 
We would content ourselves here by taking $-1 < \alpha < 0$ and $a < 0$. Then, a similar analysis gives
$E_n \propto -n^{\alpha/(\alpha+1)}$ which tends to zero from below, just as we expect for the bound 
states. Note, however, that the masslessness assumption is dubious. The highly excited states are 
better described by massive, slowly moving particles, as the average momentum in a cycle is low.
Explicitly, assuming masslessness we get
\be
\langle p \rangle =       \frac{\int p(t) dt}{\int dt} = \frac{c^{-1} \int p(x) dx}{c^{-1} \int dx}
                  \approx \frac{n}{4 x_+} \sim n^{\alpha/(1+\alpha)} \longrightarrow 0 
\ee
as $n \rightarrow \infty$. Obviously, the mass cannot be ignored and the classical form of the kinetic 
energy should be used in those cases.  
For the Coulomb force $\alpha = -1$, this leads to $E_n \propto -n^{-2}$ as in the hydrogen atom. 
Such mesons, or positronia, which are essentially two weakly bound, slowly moving particles in a classical 
(although screened) potential, were seen in \cite{Kruczenski:2003be}. 

We now turn to explore the effects of giving a finite mass $m$ to the ``quarks'', 
and of adding angular momentum $J$ to the system. 
We will deal only with the linear confining potential $\alpha = 1$, which we rewrite as 
$V(x) = 2 T x$, with $T$ the string tension, for which (\ref{nalpha}), (\ref{Enalpha}) give
\be
\label{Env0}
E_n \approx \sqrt{\frac{\pi \hbar T}{2}} \cdot \sqrt{n}
\ee
First, we think of the angular momentum of the system as coming from the particles and not from the
string itself.  
Admittedly this is not the case, but this should still serve as a good model for proving that those 
effects are small. Moreover, this can be shown to be the non--relativistic limit of the full problem
\cite{Ida:1977uy}. Later we will look at a model where the string carries the angular momentum.

The Hamiltonian of (half) the system is now   
\be
\label{HmJv1}
H = \sqrt{p^2 + m^2 + (J/x)^2} + T x
\ee
where now $p$ is the radial momentum, $p \equiv p_x$, and $J$ is the conserved momentum conjugate to the
angular variable, $J \equiv p_\phi$.
The two parameters $m$ and $J$ should be small, where we should form only classical dimensionless 
quantities --- using $c = 1$ but not $\hbar$. For the mass, this obviously leads to 
\be
h \equiv \frac{m}{E} \ll 1
\ee 
where $E$ is the total energy, while for the angular momentum we should have
\be
g \equiv \frac{J T}{E^2} \ll 1
\ee
This is equivalent to the demand that the total action $S$ given in (\ref{En0}) is much larger than $J$.

The main tool in estimating the corrections to (\ref{En0}) is the theorem that for an integral depending
on a parameter through the integrand and integration limits,
\be
I(\epsilon) \equiv \int_{a(\epsilon)}^{b(\epsilon)} k(x ; \epsilon) dx
\ee
where we think of $\epsilon$ as a small number, one has
\be
\label{Iprime}
I'(0) = b'(0) k(b(0), 0) - a'(0) k(a(0), 0) + \int_{a(0)}^{b(0)} \partial_\epsilon k(x ; 0) dx
\ee
In our case, the integrand $k$ is
\be
p(x) = \sqrt{(E - T x)^2 - m^2 - (J/x)^2} 
\ee
which can be written in terms of the dimensionless quantity $y \equiv T x / E$ as
\be
p(y ; h , g) = E \sqrt{(1 - y)^2 - h^2 - (g/y)^2}
\ee
so the two small parameters $h,g$ play the role of $\epsilon$.
Our task is facilitated by the fact that, by definition, $p(x_\pm(\epsilon) ; \epsilon) = 0$ as $x_\pm$
are the turning points, so the first two terms in (\ref{Iprime}) generically vanish and we are left only
with the integral. However, more than one differentiation might be needed, and divergent expressions
in the two $\epsilon$'s might be encountered. Omitting the technical details, we find, in accordance with
(\ref{Env0}), that
\be
\label{SmJv1}
\pi \hbar n \approx S \equiv 4 \int_{x_-}^{x_+} p(x) dx 
            \approx 4 \cdot \frac{E^2}{2 T} 
                    \left(1 + h^2 \log h - \pi g + \mbox{higher order corrections}\right)
\ee    
and therefore finally that
\be
\label{Env1}
E_n = \sqrt{\frac{\pi \hbar T}{2}} \cdot \sqrt{n} \cdot 
      \left( 1 - \frac{m^2}{\pi \hbar T} \cdot \frac{\log n}{n} 
               - \frac{2 J}{\hbar} \cdot \frac{1}{n} 
               + \mbox{higher order corrections} \right)
\ee

If, instead of (\ref{HmJv1}), we stipulate that the angular momentum arises from the string, which behaves 
as a slowly rotating, classical straight rod, we have
\be
\label{HmJv2}
H = \sqrt{p^2 + m^2} + T x + \frac{3}{2} \frac{J^2}{T x^3}
\ee
Similar analysis reveals now that the leading relative correction in $g$ is now of the order $g^{2/3}$
(instead of the linear behaviour seen in the parenthesis of (\ref{SmJv1})), so that, instead of 
(\ref{Env1}) we have
\be
\label{Env2}
E_n = \sqrt{\frac{\pi \hbar T}{2}} \cdot \sqrt{n} \cdot 
      \left( 1 - \frac{m^2}{\pi \hbar T} \cdot \frac{\log n}{n} 
               - O\left(\frac{J}{\hbar n}\right)^{2/3}  
               + \mbox{higher order corrections} \right)
\ee

In any case, we see that the results tend to resemble well the simple result (\ref{Env0}) for $h,g \ll 1$.
The exact treatment of the system is more involved than the two models presented, and is not pursued, 
but the qualitative picture would be the same.

The R--charge $J_R$ is represented by angular momentum of the string in the extra directions of the
probe brane, those that do not constitute the Minkowski space of the field theory. In order to explore
the effects of non--zero $J_R$ we should take into account the non--trivial profile of the brane in
those directions. However, it is clear that for small $J_R$, compared to $E$, we can introduce a third 
small parameter $g_R$ and expand in it, and that the limit of $g_R \rightarrow 0$ is smooth and gives 
results similar to the above.

Obviously, all this would still be true even if we had given different masses to the two ``quarks''.

\sect{Spectrum of Mesons Dual to Brane Probes Fluctuations} 
\label{Mesons:BraneFlucts}

In this section we explore the spectrum of mesons with spin $J = 0, 1$ in theories having weakly curved
dual supergravity backgrounds. The mesons themselves are dual to the fluctuation eigenmodes of static 
brane probes placed in those backgrounds. We find that in confining and non--confining theories alike,
the spectrum is linear in the excitation number, $M_n \approx M_1 n$. This generic behaviour is totally 
insensitive to the quark anti--quark potential.

Indeed, this is to be expected, as the condition (\ref{x0}) is primarily concerned with the 
$|y| \rightarrow \infty$ region, where the radial coordinate of the background is large, the Infra--Red 
details are insignificant, and, generically, the original conformal/AdS properties are manifest. 
It is interesting to note, however, that the mass squared of the mesons, when known, is approximated by a 
quadratic expression in $n$ (including a linear and constant terms) to a high accuracy even for the 
low--lying states.  

\subsection{Mesons from the \adss Geometry} 
\label{Mesons:BraneFlucts:AdS5S5}

The spectrum of scalar and vector mesons in the non--confining theory of \cite{Karch:2002xe} was computed
exactly in \cite{Kruczenski:2003be} by looking at the worldvolume theory of the D7 brane and studying its
fluctuations. All the results are of the form 
\be
\label{AdS5S5:M}
M \approx 2 m_q \sqrt{\frac{\pi}{g_s N}} (n + J_R)
\ee
where $n$ is the excitation number and $J_R$ the R--charge of the meson. Despite the exact solution, we 
wish to readdress the problem, at least for the simplest case, using our general approach, in order to
verify our method and to compare the results with those of other scenarios, which will be confining.

The D7 brane worldvolume consists of the four--dimensional Minkowski space of the field theory, a 
transverse radial coordinate we shall denote by $y$,\footnote{Our $y$ is $\rho$ in the notations of 
\cite{Kruczenski:2003be}. Confusingly, those authors use $y$ for a different, reparameterized, coordinate.}
and a three--sphere. One can separate variables in the wavefunction and write it as the product, 
respectively, of a plane wave, a radial function $\chi(y)$ and a spherical harmonic with ``spin'' $J_R$, 
which is an R--charge from the field theory point of view. The Laplacian of the Minkowski plane wave gives
the meson's mass squared, while the second Casimir eigenvalue of the spherical harmonic is $J_R (J_R + 2)$.
For the simplest scalar mesons, the resulting equation, given in (3.6) of \cite{Kruczenski:2003be}, is
\be
-\frac{(1 + y^2)^2}{y^3} \partial_y \left(y^3 \partial_y \chi(y)\right) + 
 J_R (J_R +2) \frac{(1 + y^2)^2}{y^2} \chi(y) 
= \lambda \chi(y) 
\ee 
where 
\be
\label{lambdabarM}
\lambda = \bar{M}^2
\ee
and $\bar{M}$ is the rescaled, dimensionless mass, with
\be
\label{MbarM}
M = m_q \sqrt{\frac{\pi}{g_s N}} \bar{M}
\ee
We therefore have, in the notations of subsection \ref{WKB:Laplacian}, that
\be
\Gamma(y) = \frac{y^3}{(1 + y^2)^2}         \;\; ; \;\;
\Sigma(y) = \frac{1}{1 + y^2}               \;\; ; \;\;
\Xi(y)    = \frac{y^{3/2}}{(1 + y^2)^{1/2}} \;\; ; \;\;
A(y)      = J_R (J_R + 2) \frac{(1 + y^2)^2}{y^2}
\ee
so  
\be
x(y) = \int_0^y \Sigma(\tilde{y}) d\tilde{y} = \int_0^y \frac{d\tilde{y}}{1 + \tilde{y}^2} = \arctan(y)
\ee
and $x_0 = \lim_{y\rightarrow \infty} x(y) = \pi/2$. Consequently,
\be
\Gamma(x) = \sin^3(x) \cos(x)            \;\; ; \;\;
\Sigma(x) = \cos^2(x)                    \;\; ; \;\;
\Xi(x)    = \sin^{3/2}(x) \cos^{-1/2}(x) 
\ee
and 
\be
A(x)      = J_R (J_R + 2) \sin^{-2}(x) \cos^{-2}(x)
\ee 

Taking $J_R = 0$ we find from (\ref{VxXiA})
\be
V(x) = \frac{\Xi''(x)}{\Xi(x)} = \frac{1}{4} \left(2 + 3 \tan^2(x) + 3 \cot^2(x)\right)
\ee
So that we have $V(x)$ diverging to infinity at $x = x_0 = \pi/2$. Somewhat surprisingly, we find
that the potential also diverges at $x = 0$. We therefore have effectively a square well, but with width 
$x_0$ instead of $2 x_0$. Taking this into account in (\ref{Mn}), (\ref{lambdabarM}) gives 
$\bar{M}_n \approx 2 n$, which together with (\ref{MbarM}) exactly reproduces (\ref{AdS5S5:M}). 

Taking $J_R \neq 0$, $V(x)$ changes but still has the same square well behaviour, with the same width, 
so the behaviour in $n$ is the same.
Here we have, however, a certain puzzle regarding the behaviour of the mass as a joint function of
$n$ and $J_R$. As $A(x) \propto J_R (J_R + 2) \approx J_R^2$, it na\"{\i}vely seems 
that the ground energy is raised by $A(\pi/4) \approx 4 J_R^2$, so $\lambda = 4(n^2 + J_R^2)$ and 
therefore $M \sim \sqrt{n^2 + J_R^2}$ instead of $M \sim n + J_R$.
However, we have not explored the $1/n$ corrections to the WKB result; if such a correction yields
a contribution of the form $8 n J_R$ to the eigenvalue, we will indeed have
$\lambda = 4(n^2 + 2 n J_R + J_R^2) = 4 (n + J_R)^2$ and $M \sim n + J_R$ as needed.

The other scalar mesons and the vector meson behave in a very similar fashion. 

\subsection{Mesons from the Klebanov--Strassler Geometry}

In \cite{Sakai:2003wu}, a similar analysis was carried out 
for the Klebanov--Strassler confining background \cite{KleStr}, where massless quarks have been added 
using D7 probes. 
The eigenfunction equations for the masses squared of the low spin mesons are of the
form described at subsection \ref{WKB:Laplacian}. The vector meson case is described in  
equation (4.15) of \cite{Sakai:2003wu}.
As $\Sigma(y)$ behaves\footnote{Their coordinate $\tau$ is playing the role of our $y$; 
our $\Sigma(y)$ corresponds to their $\sqrt{(K \sinh \tau)' I(\tau)}$,
and our $\Gamma(y)$ to their $\sqrt{\gamma}$.}, for large $y$, as 
$\alpha y^{1/2} e^{-y/3}$, where $\alpha$ is some known number, the new coordinate $x$ behaves as 
$x \sim \mbox{\it const} - 3\alpha y^{1/2} e^{-y/3}$ where $\mbox{\it const}$ is arbitrary. 
Near the origin $y = 0$, all the relevant functions, including $\Sigma(y)$, behave smoothly and nicely and
do not diverge. As the range of $y$ is $0 \le y < \infty$ and the boundary conditions
on the wavefunction at $y = 0$ is that either it or its derivative vanishes, we can continue the range
to $-\infty < y < \infty$ and have a symmetric potential (and therefore a symmetric or an antisymmetric
wavefunction). This can be translated into a finite range of $x$, symmetric around $0$, by choosing 
the $\mbox{\it const}$ appropriately, $-x_0 < x < x_0$ where $x_0 = \int_0^\infty \Sigma(y) dy$ is finite, 
as in (\ref{x0}). 
The function $\Xi(y)$ behaves as $y^{1/2} e^{y/6}$ for large $y$, and therefore 
\be
\label{KS:Xix}
\Xi(x) \approx \left(- \log (x_0 - |x|)\right)^b \left(x_0 - |x|\right)^a
\ee
with $a = -\half, b = \frac{3}{4}$, so the potential behaves as
\be
\label{KS:Vx}
V(x) \approx \left(x_0 - |x|\right)^{-2} \left( 1 + O\left(\frac{1}{\log(x_0 - |x|)}\right)\right)
\ee 
at $|x| \lesssim x_0$.
We therefore have essentially a square well, and the eigenvalue behave as $\lambda \propto n^2$ at high 
excitations. As the eigenvalues are the masses squared, we indeed get for the vector meson masses 
approximately $M_n \propto n$, which is identical to the non--confining result when $J_R = 0$. 

The analysis for the (pseudo)scalar cases is very similar. Indeed, $\Sigma(y)$ is the same in all cases,
so $x_0$ is the same as in the vector meson case. We still have also (\ref{KS:Xix}), although the values 
of $a,b$ change from case to case, and therefore (\ref{KS:Vx}) also still holds.
 
\subsection{Mesons from Witten's Confining Geometry}

Witten \cite{Witten:1998zw} has introduced a background dual to a four--dimensional confining gauge theory 
by compactifying D4 branes on a spatial circle having appropriate periodicity conditions.
The construction is similar to the three--dimensional case reviewed in appendix \ref{Appendix:YM3}. 
In \cite{Kruczenski:2003uq}, probe D6 branes were added to this background to introduce dynamic quarks,
in the spirit of \cite{Karch:2002sh}. The D6 branes span the four--dimensional Minkowski space of the 
gauge theory, and three transverse directions; those are specified by a radial coordinate, which we denote
by $y$,\footnote{$\lambda$ is used instead of $y$ in the notations of \cite{Kruczenski:2003uq}.} 
and a two--sphere. The classical brane solution is characterized by a function $r_{vac}(y)$ measuring,
in a sense, the distance of the D6 brane from the D4 branes; 
$\rho_{vac}(y) \equiv \sqrt{r_{vac}^2(y) + y^2}$ is also used. 
The second order differential equation of motion fixing $r_{vac}(y)$ is given in (2.17) of 
\cite{Kruczenski:2003uq}; it is quite complicated and does not seem to admit an analytical solution.
Nevertheless, the solution is parameterized by the asymptotic value 
$r_\infty$ of $r_{vac}(y)$ as $y$ tends to infinity; $r_\infty$ essentially determines linearly
the mass of the dynamic quarks. 
Moreover, an approximate solution is not hard to find \cite{Kruczenski:2003uq}:
\be
r_{vac}(y) \approx r_\infty + \frac{c}{\sqrt{y^2 + r_\infty^2}}
\ee 
where, for large $r_\infty$, one has $c \approx \frac{1}{2 r_\infty}$. 

In the same paper, an analysis of the low spin mesons is given through the quadratic action of the brane 
probe fluctuations. The equations of motion for two modes, denoted by $\delta \phi$ and $\delta r$, are 
given there in equations (3.4), (3.5). We will look only at the $\delta \phi$ mode for the case where it 
has no angular momentum on the aforementioned two--sphere, and where $r_\infty \neq 0$. Then, $\delta \phi$
is a function of $y$ only, and the eigenvalue problem for the meson mass squared is of the type 
(\ref{SchrLap}). It is a simple matter to find the behaviour of the various functions defined above at the 
extremities of the range of $y$. 

For small values of $y$,
\be
y \rightarrow 0      \;\; : \;\; \Gamma(y) \approx c_1 \, y^2 + c_2 \, y^4 \;\; ; 
                            \;\; \Sigma(y) \approx c_3        + c_4 \, y^2 \;\; ; 
                            \;\; \Xi(y)    \approx c_5 \, y   + c_6 \, y^3
\ee
where the $c_i$ are known but immaterial non--zero constants. In particular, 
$x(y) = \int_0^y \Sigma(\tilde{y}) d\tilde{y} \approx c_3 y + \frac{c_4}{3} y^3$ there, so 
$\Xi(x) = \Xi(y(x)) = c_7 x + c_8 x^3$; therefore $V(x) \equiv \frac{\Xi''(x)}{\Xi(x)} \approx c_9$ neither
vanishes nor diverges there, and we can safely continue it in a symmetric fashion to 
$-\infty < x < \infty$.  

For large values of $y$,
\be
y \rightarrow \infty \;\; : \;\; \Gamma(y) \approx r_\infty^2 \, y^{-1} \;\; ; 
                            \;\; \Sigma(y) \approx y^{-3/2}          \;\; ; 
                            \;\; \Xi(y)    \approx r_\infty \, y^{1/4}
\ee
so 
$x(y) = \int_0^y \Sigma(\tilde{y}) d\tilde{y} \approx x_0 - \int_y^\infty \tilde{y}^{-3/2} = 
x_0 - 2 y^{-1/2}$, where by the preceding discussion, $x_0$ is finite. We find, then, that
for $|x| \lesssim x_0$, one has $y \approx 4 (x_0 - |x|)^{-2}$ and therefore 
$\Xi(x) \approx \sqrt{2} r_\infty (x_0 - |x|)^{-1/2}$. Again we find an infinite potential well,
$V(x) \equiv \frac{\Xi''(x)}{\Xi(x)} \approx \frac{3}{4} (x_0 - |x|)^{-2}$, eigenvalues quadratic in the 
excitation number, and therefore meson masses linear in it.

\subsection{Mesons from the Constable--Myers Geometry}

The Constable--Myers background \cite{Constable:1999ch} is confining, albeit singular in the Infra--Red.
Luckily, D7 probe branes can be safely immersed in the background, as they are repelled from the 
singularity \cite{Babington:2003vm, Babington:2003up}. After the probe brane profile is determined, 
its fluctuations can be investigated \cite{Babington:2003vm, Babington:2003up, Evans:2004ia}; 
the equations for the fluctuations giving rise to the pseudoscalar
and vector mesons both have essentially the form (neglecting some mild factors)\footnote{In 
\cite{Evans:2004ia}, $\rho$ is used instead of $y$, and $f(\rho), g(\rho)$ replace $\chi(y)$ in their 
equations (32), (56) for the pseudoscalar and vector cases, respectively.}
\be
- {\cal G}^{-1}(y) H^{-1}(y) \frac{d}{dy} \left[{\cal G}(y) \frac{d}{dy} \chi(y) \right] = \lambda \chi(y)
\ee

Near $y = 0$, one has ${\cal G}(y) \approx y^3$ and $H(y) \approx \mbox{\it const} \neq 0$, so that
\be
y \rightarrow 0      \;\; : \;\; \Gamma(y) \sim y^3 \;\; ; 
                            \;\; \Sigma(y) \sim 1   \;\; ; 
                            \;\; \Xi(y)    \sim y^{3/2}
\ee   
and $\int_0 \Sigma(y) dy$ converges. 

For large $y$, the behaviour of ${\cal G}(y)$ stays the same, while $H(y) \sim y^{-4}$, and therefore
\be
y \rightarrow \infty \;\; : \;\; \Gamma(y) \sim y^{-1} \;\; ; 
                            \;\; \Sigma(y) \sim y^{-2} \;\; ; 
                            \;\; \Xi(y)    \sim y^{1/2}
\ee
so $x(y) \sim x_0 - y^{-1}$ for finite $x_0$, hence $\Xi(x) \sim (x_0 - |x|)^{-1/2}$ and again 
$V(x) \sim +(x_0 - |x|)^{-2}$, giving rise to $M_n^2 = \lambda \sim n^2$. Sure enough, the numerical 
values for the vector meson masses \cite{Evans:2004ia} follow the $M_n \propto n$ result.

\subsection{Mesons from the Maldacena--N\'u\~nez Geometry}

Here we briefly deal with the $J \le 1$ meson analysis \cite{Nunez:2003cf} in the Maldacena--N\'u\~nez 
confining background\footnote{We need to use the D5 formulation of the background, which is $S$ dual to 
the NS5 formulation.} \cite{Maldacena:2000yy}. 
The radial coordinate of the background is 
$0 \le y < \infty$,\footnote{The authors of \cite{Nunez:2003cf} use $r$ instead of $y$ as the radial 
coordinate.} and the probe brane is placed such that its nearest point to the origin is at $y = y_*$.
The quadratic Lagrangian of brane fluctuations is given in (7.5) of \cite{Nunez:2003cf}, and the 
equation of motion is easily derived. A redefinition of the field is needed in order for the equation 
to take the form (\ref{SchrLap}), where $\lambda = M^2$ directly multiplies the eigenfunction; 
when this is done, one finds that
\be
\Sigma^2(y) = e^{-\phi} \left( y \, \tanh y \, \cos \theta_0(y) \right)^{-1/2} 
                        \left(1 + y \, \coth y + \tan^2 \theta_0(y) \right)
\ee  
where $y_* \le y < \infty$, the dilaton $\phi$ takes some non--zero value at $y = y_*$ and behaves as 
$e^{- 2 \phi} \sim y^{1/2} e^{-2 y}$ as $y \rightarrow \infty$, and where 
$\sin \theta_0(y) \equiv \sinh y_*/\sinh y$.

At $y \rightarrow y_*$, the behaviour of all the ingredients is tame, apart from 
$\cos \theta_0(y) \sim (y - y_*)^{1/2}$, leading to the divergence $\Sigma(y) \sim (y - y_*)^{-5/8}$.
Still, $\int_{y_*}^{\hat{y}} \Sigma(y) \, dy \sim (\hat{y} - y_*)^{3/8}$ does not diverge. 
As $y \rightarrow \infty$, the exponential fall--off of $e^{-\phi}$ and of $\sin \theta_0(y)$ ensures that 
$\Sigma(y) \sim y^{3/8} e^{-3y/2}$ and therefore that $\int_{\hat{y}}^\infty \Sigma(y) \, dy$ converges.
Consequently, $\int_{y_*}^\infty \Sigma(y) \, dy < \infty$, and, as we have argued in subsection 
\ref{WKB:Laplacian}, the spectrum behaves as $M_n \propto n$.

In fact, a numerical analysis in \cite{Nunez:2003cf} reveals that 
$M \approx \sqrt{M_1^2 n^2 + \tilde{M}_1^2 J_R^2}$ for appropriate $M_1, \tilde{M}_1$. 
This is what we would have expected generically, but note that we have seen in subsection 
\ref{Mesons:BraneFlucts:AdS5S5} that the behaviour in the case of the \adss background, dual to the 
conformal SYM theory, is different.

\sect{Discussion}

We have argued that the spectrum of highly excited mesons with the same quantum charges 
(spin $J$ and R--charge $J_R$) 
behaves as $M_n \propto \sqrt{n}$ in confining field theories and in the open string picture of
mesons in the string theory duals of such field theories. 

We have further argued that the manageable (i.e.\ weakly curved) duals, confining or not, generically 
have an extra sector of (light) mesons with $J \le 1$, coming from the brane probe fluctuations, 
behaving as $M_n \propto n$. The insensitiveness of this result to the Infra--Red behaviour of the field
theory calls for a deeper understanding. 

In order to deal with the quantization of the open string living in the dual background and ending on the 
brane, we have proposed that we might assume that the string, if frozen at any instant, is in a state of 
minimal potential energy, that is, obeying the classical static equations of motion. Then, the Hamiltonian
of the system, containing also the kinetic energy, can be written in terms of the positions and momenta  
of the string endpoints. A general expression for the Hamiltonian can be written down, but is highly
complicated and its quantization seems next to hopeless.

In certain cases, most notably when confinement occurs, we argued that the ends might be simply quantized 
as particles in the potential given by the classical string solution. 
This suggestion obviously needs elaboration. Note, for example, that we essentially had to deal 
with the string potential energy but not with its kinetic energy, as the strings we considered had 
excitation numbers $n \rightarrow \infty$ but fixed spin, and the rotational energy of the string was not
the leading contribution. It would be interesting to go further, perhaps even to cases were the projection
of the classical string solution on the field theory space is not straight.  

We have treated the particles at the ends of the string as spinless and chargeless. It would be interesting
to explore the effects of their more exact description, although those effects should be subleading
in the confining cases, at least. 

In the picture of mesons as strings suspended from probe branes, much work can be done in combining radial
excitations (with quantum numbers $n$ and $n_R$) and the conserved charges (angular momentum $J$ 
and R--charge $J_R$). The latter is especially interesting, as the probe brane 
has a non--trivial profile in the extra directions giving rise to the R--charge.
The spectrum as a function of all those four parameters will reveal a lot of the Physics of the system.

It would also be interesting to explore further the case of non--confining dual string backgrounds.
We have dealt with the simplest case of \adss in the introduction and in subsection
\ref{Mesons:BraneFlucts:AdS5S5}. As we have seen, the string there is in a sense even non--relativistic.

We have extensively used the semiclassical, or WKB, approximation. It would be, of course, interesting 
to go beyond it. 

Even a higher order WKB approximation might be interesting. We remind the reader of
the puzzle presented in subsection \ref{Mesons:BraneFlucts:AdS5S5} which seems to warrant such a treatment.
A general understanding of the arising of different behaviours, 
such as $M \approx M_1 n + \tilde{M}_1 J_R$ vs.\ $M \approx \sqrt{M_1^2 n^2 + \tilde{M}_1^2 J_R^2}$, 
is lacking. 
Note that the latter relation implies an attraction between mesons,
while in the former there are only marginally bound states.
For example, we might wonder whether the latter relation is connected with confining theories 
and the former with non--confining ones.
We might also wonder when is $M_1 = \tilde{M}_1$.   

It would also be extremely interesting to extend the ideas presented in this paper to the annulon framework
\cite{Gimon:2002nr, Apreda:2003gs, Kuperstein:2003yt, Bertoldi:2004rn}.

The string endpoint quantization in general, and its WKB approximation in particular, were mainly  
applied to find the meson spectrum, that is, the energy levels. It can also be used, of course, to 
find the wavefunctions. This should give us a way to explore the form factors of the mesons, or more 
accurately of the valence quarks in them --- the sea gluons (and quarks) should be described by the 
string itself.

Lastly, the methods presented in this work should also be applicable to baryons in the dual string picture.
Those baryons are a collection of open strings with one end on the probe brane and the other end on a 
wrapped brane (the baryon vertex); a better dynamic understanding of them is highly desirable.


\vspace{1cm}
\noindent
{\bf Acknowledgments}\\
We would like to thank Ofer Aharony, Dominic Brecher, Yitzhak Frishman, Kazuyuki Furuuchi, Moshe Rozali, 
Jacob Sonnenschein, Matthew Strassler, David Oaknin and Ariel Zhitnitsky for valuable discussions.
We also thank the seminar audiences at the University of British Columbia, the Perimeter Institute and 
the University of Toronto for enlightening questions and comments when this work has been presented.


\appendix

\renewcommand{\theequation}{\Alph{section}.\arabic{equation}}

\sect{The Meson Spectrum in QCD}
\label{Appendix:DispersionRelations}

We aim to show that the behaviour of the spectrum of excited mesons in QCD is $M_n \propto \sqrt{n}$,
which we rewrite as 
\be
\label{MnM1n}
M_n \approx M_1 \sqrt{n}
\ee
As previously stated, this behaviour can be found analytically in two--dimensional large $N$ QCD 
\cite{'tHooft:1974hx}.
However, using the sum rules arising from dispersion relations \cite{Shifman:bx}, it is possible to arrive 
at that result generically and easily, even if not totally without assumptions 
\cite{Zhitnitsky:1985um, Kogan:1995nd}. We review that approach in this appendix.

Let us start with two--dimensional QCD. Define the two point time ordered correlation function
\be
\widehat{\Pi}(x) = 
\left<0\right| T \left[ \bar{\psi} \psi (x) \, \bar{\psi} \psi (0) \right] \left|0\right>
\ee
where $\psi$ is the quark field, and its Fourier transform
\be
\Pi(Q^2) = \int dx \, e^{i q x} \, \widehat{\Pi}(x)
\ee
where $Q^2 = -q^2$. The asymptotic $Q^2 \rightarrow \infty$ behaviour is governed by the 
free theory, as the effective, dimensionless coupling constant is $g/Q \rightarrow 0$. 
Obviously, as a loop diagram, $\Pi(Q^2)$ diverges (see figure); renormalizing it (using 
dimensional regularization, for example), effectively subtracts $\Pi(\mu^2)$ at some renormalization 
scale $\mu$. One then easily obtains 
\be
\label{tildePiQ2:DimReg}
\widetilde{\Pi}(Q^2) \equiv \Pi(Q^2) - \Pi(\mu^2) \sim -\log (Q^2/\mu^2)
\ee      
where we are not concerned with the proportionality constant.

\begin{figure}[h!]
\begin{center}
\resizebox{0.8\textwidth}{!}{\includegraphics{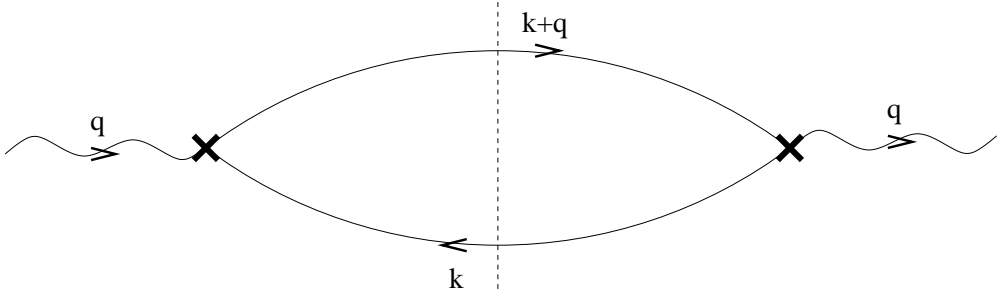}}
\end{center}
\caption{The Feynman diagram of $\Pi(Q^2)$}
\label{fig:CorrFctn}
\end{figure}

From the dispersion relations \cite{Shifman:bx}, on the other hand, one gets
\be
\Pi(Q^2)         \sim  \sum_n \frac{f_n^2}{Q^2 + M_n^2}
\ee 
and therefore
\be
\label{tildePiQ2:Dispersion}
\widetilde{\Pi}(Q^2) \sim -\sum_n \frac{f_n^2 \, (Q^2 - \mu^2)}{(Q^2 + M_n^2) (\mu^2 + M_n^2)}
\ee
where $n$ ranges over the whole spectrum of possible intermediate states $\left|n\right>$ of masses
$M_n$. 
The matrix elements are defined as
\be
f_n \equiv \left<0\right| \bar{\psi} \psi \left|n\right>
\ee
Obviously, we should only consider states with the correct quantum numbers for that channel, that is, 
scalar particles (having zero spin and positive parity). Taking into account also the pseudoscalars
only involves doubling of the density of states. 
Note also that we ignore here, as is appropriate for large $N$ theories, 
the widths of those particles.
This, of course, is not a good picture for real life $N = 3$ QCD.

We now make the reasonable assumption that $f_n$ tends to a non--zero constant as $n \rightarrow \infty$
(again, this can be analytically shown, and the constant computed, using the methods of 
\cite{'tHooft:1974hx}). We then get from (\ref{tildePiQ2:DimReg}), (\ref{tildePiQ2:Dispersion}) that,
for $Q^2 \gg \mu^2$, 
\be
\label{logQ2Sum}
-\log (Q^2/\mu^2) \sim \widetilde{\Pi}(Q^2) \sim -\sum_n \frac{Q^2}{(Q^2 + M_n^2) (\mu^2 + M_n^2)}
\ee
We can now see that the way to satisfy this is to have indeed (\ref{MnM1n}). 
Let us assume this relation.
The above sum in (\ref{logQ2Sum}) has three contributions. The low mass states with $M_n < \mu$, 
or with $n < n_\mu \equiv (\mu/M_1)^2$, give essentially a constant contribution 
$\sum_{n = 1}^{n_\mu} \frac{1}{\mu^2} =  \frac{1}{M_1^2}$. The high mass states with $M_n > Q$,
or with $n > n_Q \equiv (Q/M_1)^2$ behave likewise, as their contribution is essentially
$\sum_{n = n_Q}^{\infty} \frac{Q^2}{M_n^4} \approx \frac{1}{M_1^2}$. The main contribution is that of
masses in between, $\mu \le M_n \le Q$, or $n_\mu \le n \le n_Q$, essentially giving indeed
$\sum_{n = n_\mu}^{n_Q} \frac{1}{M_n^2} \approx \log (Q^2/\mu^2)$.
It is also easy to see that the argument goes both ways --- the only way to satisfy (\ref{logQ2Sum}) is 
essentially by having (\ref{MnM1n}).

The four--dimensional case is quite similar. In this case it is the easiest to look at the rho channel
of spin one mesons, by defining, as in the electric charge renormalization computation, 
\be
\widehat{\Pi}^{\mu \nu}(x) = 
\left<0\right| 
     T \left[ \bar{\psi} \Gamma^\mu \psi (x) \, \bar{\psi} \Gamma^\nu \psi (0) \right] 
\left|0\right>
\ee
Then, from the Ward identity (i.e.\ gauge invariance), the Fourier transform has the structure
\be
\Pi^{\mu \nu}(Q^2) = (q^2 \eta^{\mu \nu} - q^\mu q^\nu) \, \Pi(Q^2)
\ee
where the Physics lies in $\Pi(Q^2)$. Again the large $Q^2$ limit is governed by the free theory,
now because the full theory is asymptotically free, and $\Pi(Q^2)$ can be shown to behave, after
regularization, as $\log(Q^2/\mu^2)$. Following the arguments of the two--dimensional case, again
assuming that the $f_n$ tend to a non--zero constant, produces the same result.

\sect{The Non--Supersymmetric YM$_3$ Confining Case}
\label{Appendix:YM3}
 
A stack of $N$ coincident D3 branes, $N \gg 1$, gives rise to a supergravity background, 
whose decoupling limit is dual to ${\mathcal N} = 4$ $SU(N)$ four--dimensional Yang--Mills theory 
(SYM$_4$) \cite{Mal1}. 
When one direction, call it $x_3$, is compactified to a circle, and appropriate boundary conditions for 
that direction are chosen, supersymmetry is completely broken, resulting with the ``pure'' 
three--dimensional theory YM$_3$, which is supposed to be confining \cite{Witten:1998zw, BISY2, GrOl1} 
(there are spurious 
Kaluza--Klein excitations with the same energy scale as the glueball masses, though). 
The corresponding supergravity dual can be found from the non--extremal D3 case via a double Wick 
rotation taking 
$t \leftrightarrow x_3$, giving
\be
ds^2 = \alpha' \left[ \frac{U^2}{R^2} \left( -dt^2 + dx_1^2 + dx_2^2 \right) + 
                      \frac{U^2}{R^2} F(U) dx_3^2 + \frac{R^2}{U^2} F(U)^{-1} dU^2 + 
                      R^2 d\Omega_5^2                                                \right] 
\ee
where $F(U) = 1 - \frac{U_T^4}{U^4}$. $U_T$ is a parameter with the dimension of energy, and at $U = U_T$
there is a horizon.
The radius of the $\mbox{AdS}_5$ and $S^5$ in string units, $R$, obeys $R^4 = 4 \pi g_s N$.

Let us put an additional D3 brane at $U_B \gg U_T$ (it has a negligible backreaction on the background).
An open string hanging from this brane and reaching for lower values of $U$ is representing the Wilson line
in the dual field theory; together with the endpoints, it resembles a meson. 
The Wilson line has length $L$, with endpoints at, say, 
$x \equiv x_1 = \pm L/2, \, x_2 = 0$, so $U(\pm L/2) = U_B$, 
and the function $U(x)$ minimizes the string mass.
This mass, or energy, is given as 
$
E = \frac{1}{2 \pi}               \int_{-L/2}^{+L/2} {\mathcal L}  \,             dx = 
    \frac{1}{2 \pi} \cdot 2 \cdot \int_{U_0} ^{U_B}  {\mathcal L} \; (U'(x))^{-1} dU
$
where, from the Nambu--Goto action,
$
{\mathcal L} = \sqrt{\frac{U(x)^4}{R^4} + \frac{U'(x)^2}{1 - U_T^4/U(x)^4}}
$,
and $U_0$ is the minimal value of $U$ obtained by the string at its center.

It can be shown \cite{mine2} that for large values of $L$, one has 
$U_0 \gtrsim U_T$; in fact, $U_0 - U_T$ is exponentially small in $L$.
Moreover, the configuration looks like a bathtub \cite{mine3} --- there are two almost ``vertical'' parts 
of the string near the endpoints, going from $U = U_B$ to $U \approx U_0$ with little change in $x$. 
More precisely, the change in $x$ is growing with $L$, but only logaritmically, or very slowly. 
Then, there is the middle almost ``horizontal'' part at $U \approx U_0$ going from 
$x \approx -L/2$ to $x \approx +L/2$. 
The configuration is smooth, of course, but the connecting parts are small. Therefore, the whole 
configuration is similar to the na\"{\i}ve case discussed above, where the ``vertical'' parts play
the part of the quarks, and the ``horizontal'' part that of the string in the flat Minkowski space.
It is then natural that the effective string tension is given by the one for a string lying 
horizontally at the minimum value $U_T$ of $U$, i.e.\ 
$T \approx  \frac{U_T^2}{2 \pi R^2}$, and that the mass of the string can 
be approximated by the one for a vertical one going from the brane to the horizon,
$
m \approx \frac{1}{2 \pi} \int_{U_T}^{U_B} \frac{dU}{\sqrt{1 - (U_T^4/U^4)}} = \frac{U_B}{2 \pi} + O(U_T)
$. It can be rigorously shown that this approximation is a good one for the total energy of the 
configuration. The approximation is so good such that the quantum corrections, resembling the L\"uscher 
term (see appendix \ref{Appendix:Luscher}), are bigger than the classical ones \cite{GrOl1, mine2, mine3}.
All the relevant coefficients are explicitly known. 

Moreover, when the string endpoints move, the ``vertical'' parts acquire kinetic energy, simulating 
the quarks. In contrast, there is no transverse motion of the ``horizontal'' part, so it supplies only
the linear potential energy. 

We therefore suggest that the na\"{\i}ve model is applicable to this case with the aforementioned 
identifications. 

This situation is essentially the same for other confining backgrounds such as the YM$_4$ one. 
In fact, this is the generic situation. In non--generic ones, the details might change and the 
approximations are less spectacular, but they are still good enough for our purposes.  

For an exploration of the quadratic quantum corrections to the Regge trajectories in confining theories, 
see \cite{PandoZayas:2003yb}.

\sect{The L\"uscher Term in Flat Space}
\label{Appendix:Luscher}
 
In this appendix we review the so--called L\"uscher term \cite{Luscher}, which is the contribution of the
quantum fluctuations, at quadratic order, to the energy of an open string held at its two endpoints
and kept at a fixed length $L$. In fact, this term is nothing more than the Casimir energy for a one
dimensional quantum field theory. This computation sheds light on the quantum 
fluctuations of the string solutions we are studying. The Nambu--Goto formulation of the action is simpler
to use in this case, but, of course, the Polyakov formulation gives the same results.

The classical solution is simply $X^\mu = (t, x) = (\tau, \sigma)$, where $0 \le \sigma \le L$,
and with all transverse directions vanishing. 
We now include the fluctuations of a representative transverse direction $y$: 
$X^\mu = (t, x, y) = (\tau, \sigma, y(\tau,\sigma))$. Working with the ``mostly positive'' convention 
for the flat Minkowski metric, we get 
$
h_{\alpha \beta} \equiv \partial_\alpha X^\mu \partial_\beta X^\nu \eta_{\mu \nu} = 
\begin{pmatrix}
-1 + \dot{y}^2  &     \dot{y} y' \\
     \dot{y} y' & 1 + y'^2
\end{pmatrix}
$,
so $\det h = -1 + \dot{y}^2 - y'^2$. 
Therefore, the Lagrangian density, to the second order in the fluctuation $y$, is  
$
{\mathcal L} = -T \sqrt{|\det h|} =  -T \left(1 - \half \left(\dot{y}^2 - y'^2\right)\right)
$.
The Hamiltonian density, in terms of $y$ and its conjugate momentum
$
\pi_y \equiv \frac{\partial {\mathcal L}}{\partial \dot{y}} = T \dot{y} 
$
is 
$
{\mathcal H} \equiv \pi_y \dot{y} - {\mathcal L} = T + \half \left(\frac{1}{T} \pi_y^2 + T y'^2\right) 
$.
The string energy is therefore 
$
E = \int {\mathcal H} d\sigma 
  =  T L + T \int_0^L \left( \half \dot{y}^2 + \half y'^2\right) d\sigma
$. 
The first term is the classical energy coming from the length of the string, while the second one is
the quadratic fluctuation contribution $E^{(2)}$. If the string is held fixed at its endpoints, the
boundary condition on $y(\tau,\sigma)$ is of the Dirichlet kind, $y(\tau,0) = y(\tau,L) = 0$. We can 
then Fourier expand $y$ as 
$y(\tau,\sigma) = \sum_{n = 1}^{\infty} a_n(\tau) \frac{1}{\sqrt{L T/2}} \sin(\frac{\pi \sigma n}{L})$
and get $E^{(2)} = \sum_{n = 1}^{\infty} \half \dot{a}_n^2 + \half \omega_n^2 a_n^2$ where the 
frequency of the $n$-th mode is $\omega_n = \frac{\pi n}{L}$. The Casimir energy is the sum of the 
vacuum energies for each mode, when this expression of the energy is first quantized,
$
E^{(2)} = \sum_{n = 1}^{\infty} \half \hbar \omega_n 
        = \frac{\pi}{2} \cdot \frac{1}{L} \cdot \sum_{n = 1}^{\infty} n
$. 
We regularize this expression using the Riemann zeta function, 
$\sum_{n = 1}^{\infty} n \equiv \zeta(-1) = -\frac{1}{12}$. Remembering also that a string in $D$
spacetime dimensions has $D-2$ transverse directions, each of them behaving as our $y$, we finally 
get the L\"uscher term 
$E^{(2)} = - \frac{(D-2) \pi}{24} \cdot \frac{1}{L}$.
Notice that this term is Coulombic, having no dependence on the dimensionful string tension $T$.
It is also always attractive.
We have taken only the first term in the semiclassical expansion (the expansion in powers of 
$\hbar$); that is valid if this term is much smaller than the zeroth order one, 
$\frac{1}{L} \ll T L$, or equivalently if the string is held at a much larger length than its natural 
one, $L \gg \frac{1}{\sqrt{T}}$.  
It is easy to see that for Neumann boundary conditions, $y'(\tau,0) = y'(\tau,L) = 0$, one gets the same 
results, while for mixed Dirichlet--Neumann boundary conditions one encounters 
$\sum_{n = 0}^{\infty} (n + \half) \equiv +\frac{1}{24}$ giving 
$E^{(2)} = + \frac{(D-2) \pi}{48} \cdot \frac{1}{L}$ and a repulsive force.
For boundary conditions interpolating between those of Dirichlet and Neumann, the so--called Robin 
boundary conditions, the results are similar, as $T$ effectively drops out, and the only dimensionful
parameter remaining is $L$. 
For the superstring, the Fermionic degrees of freedom give the same contribution with the opposite 
sign, and the L\"uscher term vanishes. In backgrounds other than the flat Minkowski one this is
not necessarily so \cite{mine3}, but the validity of the expansion is certainly not worse than for the 
Bosonic string we have presented.

The L\"uscher term is the first in the quantum expansion, having the parameter 
$\frac{1/L}{T L} = \frac{\alpha'}{L^2}$. As we saw, the corresponding parameter for long strings in
\adss space is $\frac{\alpha'}{R^2} \sim \frac{1}{\sqrt{\lambda}}$. 

For a string kept at a fixed length due to rotation and the centripetal force, one can still show 
that in a regime of large angular momentum, the quantum fluctuations contribute as in the case described 
above; therefore, a long string is again enough to ensure classicality. This is discussed in the next 
appendix.   

\sect{The Stationary Rotating String in Flat Space}
\label{Appendix:RotatingString}

The well known solution of a straight open string, of tension $T$, rotating with a fixed angular frequency
$\omega$ in the $x$--$y$ plane of a flat Minkowski space, is given by
\be
X^\mu(\tau,\sigma) = (t, x, y) = 
                \left(\tau, \frac{1}{\omega} \sin(\omega \sigma) \cos(\omega \tau), 
                            \frac{1}{\omega} \sin(\omega \sigma) \sin(\omega \tau)\right)
\ee
The explicit form $\sin(\omega \sigma)$ of the radial function endows the worldsheet with a flat induced
metric, as it is easily seen that 
$h_{\alpha \beta} \equiv \partial_\alpha X^\mu \partial_\beta X^\nu \eta_{\mu \nu} = 
 \cos^2(\omega \sigma) \, \eta_{\alpha \beta}$.
The radial function is therefore determined in the Polyakov approach; it is still convenient in the 
Nambu--Goto approach we will adopt.

The range of the spatial coordinate of the string is given by $-\sigma_0 \le \sigma \le +\sigma_0$ with
$\sigma_0 = \pi/2 \omega$. The length of the string is therefore $L = 2/\omega$, and the free string 
endpoints move with the speed of light, 
$v_{\scriptscriptstyle \mathrm endpt} = \omega \cdot L/2 = 1$. The total energy of the string is 
$E = \pi T/\omega$ and its total angular momentum is $J = \pi T/2 \omega^2$; we therefore 
have the Regge trajectory $E^2 = J/\alpha'$ with $\alpha' = 1/2 \pi T$. Note that the string is long,
$L \gg 1/\sqrt{T}$, exactly when its energy is large, $E \gg \sqrt{T}$, the angular momentum is high,
$J \gg 1$, or the angular frequency is small, $\omega \ll \sqrt{T}$.

Adding a bosonic fluctuation in a transverse direction $z$, one easily gets  
\newline
$
h_{\alpha \beta} = 
                    \begin{pmatrix}
                     -\cos^2(\omega \sigma) + \dot{z}^2 & \dot{z} z' \\
                     \dot{z} z'                         & \cos^2(\omega \sigma) + z'^2
                    \end{pmatrix}
$
(we work in the ``mostly positive'' convention for the metric), and so the Lagrangian density becomes
\be
{\cal L} = \sqrt{-\det h_{\alpha \beta}} = \cos^2(\omega \sigma) + \half (z'^2 - \dot{z}^2) + O(z^4)
\ee
The quadratic contribution in $z(\tau, \sigma)$ is exactly of the form encountered in the L\"uscher term
of appendix \ref{Appendix:Luscher}; the space differentiation is with respect to $\sigma$, but as 
$\sigma_0$ is proportional to $L$, the one loop quantum contribution is again proportional to $1/L$.

For a fluctuation in the $x$--$y$ plane, which we might use the diffeomorphism invariance in order to 
write as
\be
X^\mu(\tau,\sigma) = \left(\tau, 
           \frac{1}{\omega} \sin(\omega \sigma) \cos(\omega (\tau + \phi(\tau, \sigma))), 
           \frac{1}{\omega} \sin(\omega \sigma) \sin(\omega (\tau + \phi(\tau, \sigma)))\right)
\ee
we similarly get, after using the boundary conditions, that the quadratic contribution to the Lagrangian
density is
\be
{\cal L}^{(2)} = \half \tan^2(\omega \sigma) \left(\phi'^2 - \dot{\phi}^2\right)
\ee
and a simple scaling argument \cite{mine3} shows again that the contribution is proportional to $1/L$.

When quarks are added to the string endpoints, the calculations are more involved. We omit the details,
but it can be verified that the picture described above is still valid for the sufficiently excited
string, that is, for the long, massive and high angular momentum one (those, again, are all equivalent).

\sect{Strings Excited along the Extra Dimensions}
\label{Appendix:x4string}

Here we further explore the spectrum of mesons dual to strings that are suspended from the probe brane.
We deal with the string excitations not along the field theory space directions, but along the extra 
dimensions of the probe brane, and show they are much more massive: $M_n \sim m_s \sqrt{n}$ where
$m_s \equiv \alpha'^{-1/2}$ is the fundamental string mass scale.

We will deal with the \adss background. However, as the deformations primarily affect the Infra--Red 
part of the background, the results will also apply to confining backgrounds, especially when 
$m_q \gg \Lambda_{QCD}$.

The D7 brane spans the worldvolume coordinates of the D3 branes, $x^0, x^1, x^2, x^3$, as well as
$x^4, x^5, x^6, x^7$. It is parallel to the D3 stack (therefore the configuration remains BPS) 
and removed from it, without loss of generality, in the $x^8$ direction, i.e.\ having $x^8 = D,\, x^9 = 0$.
We work with the convention that the extra directions $x^4, \cdots, x^9$, and therefore also the radial
coordinate $U = \sqrt{\sum_{i = 4}^9 (x^i)^2}$, have dimensions of energy; they are the usual coordinates
divided by $\alpha'$.
The probe brane spans, for $U > D$, a three--sphere in the $S^5$; 
the radius of the $S^3$ vanishes at $U = D$ and the D7 brane disappears there \cite{Karch:2002sh}.   

Let us pause briefly to discuss the energy scales of the problem. The mass of a bare quark, connecting
the probe D7 brane with the stack of D3 branes, is, as evident from the well known \adss metric,
$m_q = \int_0^D dU = D$.
We should think \cite{Mal1} of $m_q$ as fixed while taking the decoupling limit $m_s \rightarrow \infty$.
The tension of a string near the ``bottom'' of the brane at $U = D$ gives a Regge trajectory for short 
strings with typical mass scale of $m_{R.T.} = \frac{D}{R} = \frac{m_q}{\sqrt[4]{\lambda}}$ where $R$ 
is the AdS radius in string units and $\lambda$ the (large) 't Hooft coupling \cite{Kruczenski:2003be}. 
Finally, the mass gap is determined by the lightest excitations, which are, as described in the 
introduction, brane probe fluctuation modes \cite{Kruczenski:2003be} with 
$m_{gap} \sim \frac{m_q}{\sqrt{\lambda}}$. 
Therefore one has the hierarchy $m_{gap} \ll m_{R.T.} \ll m_q \ll m_s$.  

We now suspend a string from the D7 brane, such that it is extended in the radial direction of the extra 
coordinates, or, without loss of generality, in the $x^4$ direction: its
endpoints are at $x_4 = \pm x$ and therefore $U = U_B$ where
\be
\label{UB}
U_B = \sqrt{x^2 + D^2}
\ee
Let us also designate by $U_0$ the minimal value of $U$ attained by the string (at $x_4 = 0$).   
The angle $\theta$ that the string subtends on the sphere approaches $\pi$ as $x$ grows. In fact 
\cite{Mal2}, as 
\be
\frac{\theta}{2} = \int_1^{U_B/U_0} \frac{dy}{\sqrt{y^2 (y^2 - 1)}}
\ee
and $\cos \frac{\theta}{2} = \frac{D}{U_B}$, one easily gets 
\be
\frac{D}{U_B} \approx \int_{U_B/U_0}^\infty \frac{dy}{\sqrt{y^2 (y^2 - 1)}} 
              \approx \int_{U_B/U_0}^\infty \frac{dy}{y^2}                 = \frac{U_0}{U_B}
\ee
so that the minimal value of $U$ remains approximately constant as $x$ increases, $U_0 \approx D$,
and the string follows the probe brane quite closely.

We can now view the half string, which primarily extends in the $U$ direction for large $x$, 
as a quark with the mass $m(x) \approx \int_D^{U_B} dU = U_B - D$.
As $D$ is held fixed, and (\ref{UB}) holds, $U_B$ and $x$ are interchangeable, and we can further 
approximate $m(x) \approx x$. The system is well described by a model of free quarks with varying mass,
and the Hamiltonian for half the system reads
\be
H = \sqrt{p^2 + m^2(x)} \approx \sqrt{p^2 + x^2}
\ee
with $p \equiv \partial/\partial_x \approx \partial/\partial_U$, which is curiously a square root of the
classical harmonic oscillator. 
Here we are implicitly using the string units, because of the non--canonical dimension of $x$.
The classical momentum is $p(x) \approx \sqrt{E^2 - x^2}$, and following 
the WKB procedure of section \ref{WKB} yet again, we find
\be
\pi n \approx \int_{x_-}^{x_+} p(x) \, dx \approx \int_{-E}^{+E} \sqrt{E^2 - x^2} = \frac{\pi}{2} E^2
\ee
so the energy of the system behaves as $E \sim m_s \sqrt{n}$, where we have reinstituted the string mass 
scale. We indeed see that the excitations are extremely heavy, and grow as the square of the excitation
number, in marked contradistinction to the brane fluctuation results.

\sect{More about the Laplacian Spectrum}
\label{Appendix:MoreLaplacian}

There exists an alternative point of view, initiated by the classical analysis of Weyl, 
on the issue of the eigenvalues of a Laplacian, from which a generic answer is also expected.
On a flat Euclidean $d$ dimensional box of finite volume, the number of eigenvalues corresponding to 
momentum of absolute value $p$ behave as the area of a $p-1$ dimensional sphere, 
\be
\frac{d n}{d p} \propto p^{d-1}
\ee   
where the proportionality factor is essentially the volume of the box. The eigenvalue of (minus the) 
Laplacian is 
\be
\label{lambdap}
\lambda = p^2
\ee
Therefore, the density of the eigenvalue distribution is
\be
\label{rholambda}
\rho(\lambda) \equiv \frac{d n}{d \lambda} = \frac{d n}{d p^2} = (2 p)^{-1} \frac{d n}{d p} \propto p^{d-2}
 = \lambda^{-1 + d/2}
\ee
so
\be
\label{nlambda}
n \propto \lambda^{d/2}
\ee
or 
\be
\label{lambdan}
\lambda_n \sim n^{2/d}
\ee

For a curved manifold, whenever the eigenvalue is large enough and the eigenfunction oscillates on a scale 
much smaller than the radius of curvature, the effect of non--flatness is subleading and the behaviour
is essentially that of the flat case. When the manifold is non--compact and has boundaries (so its volume 
is still finite), we should assume that the extrinsic curvature of the boundary is also small. When the 
manifold is compact and without boundaries, the flat analogue is actually a torus, or the box with 
periodic boundary conditions.

When the manifold has a Lorentzian signature, the above reasoning still holds, but the interpretation of 
the Laplacian eigenvalue should be changed to the mass squared, 
\be
\label{lambdam}
\lambda = M^2
\ee

This analysis can be made rigorous, most easily perhaps, using the heat kernel technique. Defining the 
(integrated) heat kernel $K(t)$ essentially as the Laplace transform of $\rho(\lambda)$, that is, 
\be
K(t) \equiv \Tr \, e^{-t \Delta} = \sum_n e^{-\lambda_n t} \approx 
            \int \rho(\lambda) e^{-\lambda t} d \lambda
\ee  
it is possible to compute its asymptotics for small $t$ and show \cite{Gilkey} that
\be
K(t) = a_0 t^{-d/2} + O(t^{(1-d)/2})
\ee
where the first Seeley--De Witt coefficient,
\be
a_0 = (4 \pi)^{-d/2} \int_M d^d x \sqrt{|\det g|} 
\ee
is indeed essentially the volume of the manifold. Taking the inverse Laplace transform brings us back to
(\ref{rholambda}).

In the case at hand, after all the other quantum numbers (spin and R--charge) had been fixed, we are 
effectively left with the one dimensional problem (\ref{SchrLap}). Taking $d = 1$ in 
(\ref{lambdan}), (\ref{lambdam}) we get that for the large eigenvalues, that is, the highly radially 
excited mesons (with $J \le 1$), indeed $M_n \propto n$. 


\end{document}